\documentclass{hotnets2020}
\usepackage{times}
\usepackage{hyperref}
\usepackage{graphicx}
\usepackage{latexsym}
\usepackage{amsmath}
\usepackage{url}
\usepackage{comment}

\hypersetup{pdfstartview=FitH,pdfpagelayout=SinglePage}

\begin{document}



\title{The Remaining Improbable:\\Toward Verifiable Network Services}





\author{Pamela Zave, Jennifer Rexford, and John Sonchack\\
Princeton University}

\maketitle

\begin{abstract}
The trustworthiness of modern networked services is too important to 
leave to chance.  
We need to design these services with specific properties in mind, 
and verify that the properties hold.  
In this paper, we argue that a compositional network architecture, 
based on a notion of layering where each layer is its own complete 
network customized for a specific purpose, is the only plausible
approach to making network services verifiable.
Realistic examples show how to use the architecture to
reason about sophisticated 
network properties in a modular way.
We also describe a prototype in which the basic structures of the
architectural model are implemented in efficient P4 code for
programmable data planes,
then explain how this scaffolding fits into an integrated process
of specification, code generation, implementation of additional
network functions, and automated verification.
\end{abstract}

\section{Introduction}

Networks are an indispensable, mission-critical, and safety-critical
part of global infrastructure.
Network researchers and engineers now have a responsibility to work toward
networks whose services and properties are verifiable.
For this reason, the last decade has seen considerable research on
network verification.
Its success demonstrates that
automated verification applies to real networks and scales well enough
to be useful in practice \cite{netverify}.

The problems with current network verification are that it is low-level
and lacks modularity \cite{netverify2}, because 
it is focusing on packet delivery in physical networks,
and on analysis of their routing and forwarding.
We know that the Internet has many levels of virtual networks,
and that these virtual networks are usually made of {\it ad hoc}
tunnels, implemented by packet rewriting and encap/decapsulation
\cite{cacm}.
Even if an existing network verifier can handle the packet transformations,
neither the analysis nor the properties being checked
reflect any awareness of the virtual networks.
So the properties of virtual networks (with their own
virtual names and topologies) 
cannot be verified, even though they are
the networks that actually provide user services.
And the strong natural modularity that comes from layering of virtual
networks is unexploited during analysis.

Naturally there are efforts to raise the level of abstraction
of network verification,
but this is almost impossible when working from the bottom up.
For example, researchers have suggested getting higher-level
specifications by inferring them from existing artifacts.
Existing artifacts, however, are full of defects
accreted over time, as features were added and interwoven without
the benefit of foreknowledge.
If we had higher-level specifications we could clean up the defects,
but no automated process is likely to extract pure intentions from
messy artifacts.

We believe it is time to learn how to build network services that are
modular and verifiable by construction.
To do this, we begin with an architectural model 
that is precise, realistic, and formalizable,
with layering and modularity as central concepts.
\S\ref{sec:Compositional-network-architecture} gives a brief overview
of this model, explaining why its definition of layering is better
than the familiar layers in the ``classic'' Internet architecture.
In \S\ref{sec:Network-services-and-properties} we use an
example to show how network services and their required properties
can be specified with the model, even though they are built on virtual
structures as much as physical ones.
In \S\ref{sec:Modular-reasoning-within-the-architecture}, 
which is arguably the most important section in the paper, we
discuss reasoning with the architectural model.
The vast majority of properties can be verified by analyzing one separate
module at a time, leading to major improvements in efficiency.
Properties verified with completely different techniques can be
composed to guarantee overall network behavior.

Having established that the architectural model meets our needs,
our next task is to embody it directly in network components.
This means that the modular structures of the architecture will be
explicit in the implementation, so they can be exploited not just
for enhanced and verified services, 
but also for measurement, management, and
resource allocation.
To reach this goal, programmable data planes are the key enabler.
In \S\ref{sec:Prototype-implementation} we describe our work in progress
on a prototype implementation in the P4 language.
We show how simple
optimizations make it run efficiently, {\it i.e.,} consuming about
3\% of compute resources on the Barefoot Tofino.
We also explain our plans
to generate data- and control-plane code from templates and/or 
specifications.

\begin{figure}[hbt]
\begin{center}
\includegraphics[scale=0.50]{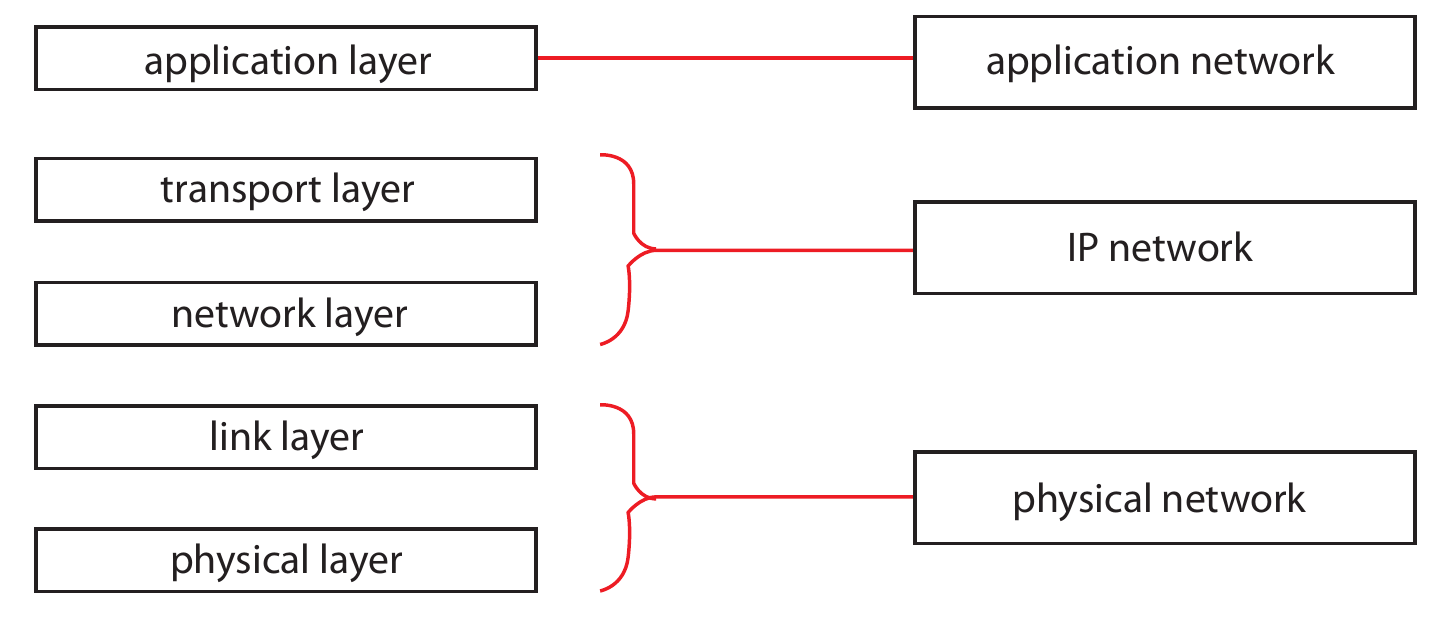}
\end{center}
\caption
{\small
{Layers in the classic Internet architecture (left),
corresponding to layered networks in compositional
network architecture.}
\normalsize}
\label{fig:correspond}
\end{figure}

There are two obvious objections to this research program.
The first is that it is a clean-slate approach, and therefore
too ambitious for credibility.
This is not true, simply
because the architectural model is drawn from close observation of today's
networks \cite{cacm}.
For this reason modules and components in our architecture should
interface easily with existing
network hardware and software, and can interoperate with
them in real deployments.

The second objection to this approach
lies in the practical implications of such intensive use of
explicit architectural 
structure and programmability in network implementations.
Especially in the short term, it may not be efficient
enough for real deployment.
We have great hopes for optimizations and the march of
chip technology, but no guarantees on them.
Despite this objection, the research is worth doing,
because model-based development is {\it the only plausible approach} to 
building networks whose services are verifiable in any meaningful sense.
Even if our prototype is never complete or efficient enough to be
used in production, all of the abstractions, properties,
and implementation mechanisms we discover can be used directly
by other researchers, and will find their way into practice.
To quote Sherlock Holmes, ``. . . when you have eliminated the
impossible, whatever remains, however improbable, must be the truth.''

The paper concludes with related work (\S\ref{sec:Related-work})
and a summary of research plans (\S\ref{sec:Research-plans}).
\S\ref{sec:Network-services-and-properties} through 
\S\ref{sec:Prototype-implementation}
present new research,
including a formal semantics for the architectural model,
based on the original ideas in \cite{cacm}.

\section{Compositional network\\architecture}
\label{sec:Compositional-network-architecture}

{\it Compositional network architecture} is based on the ideas
that a network is a module, and that network services today are
provided by rich, flexible compositions of heterogeneous network modules.
Each network/module
is a microcosm of networking, with all the basic mechanisms
including a namespace, members, links,
routing, forwarding, session protocols, and directories.
Networks are composed by {\it bridging}, with the obvious meaning.
They are also composed by {\it layering,} which means something new
and very specific.
In any network, a {\it session} is an instance of one of the services
provided by that network.
One network is layered on another network if a (virtual) 
link in the overlay
network is implemented by a session in the underlay network.

\begin{figure}[hbt]
\begin{center}
\includegraphics[scale=0.60]{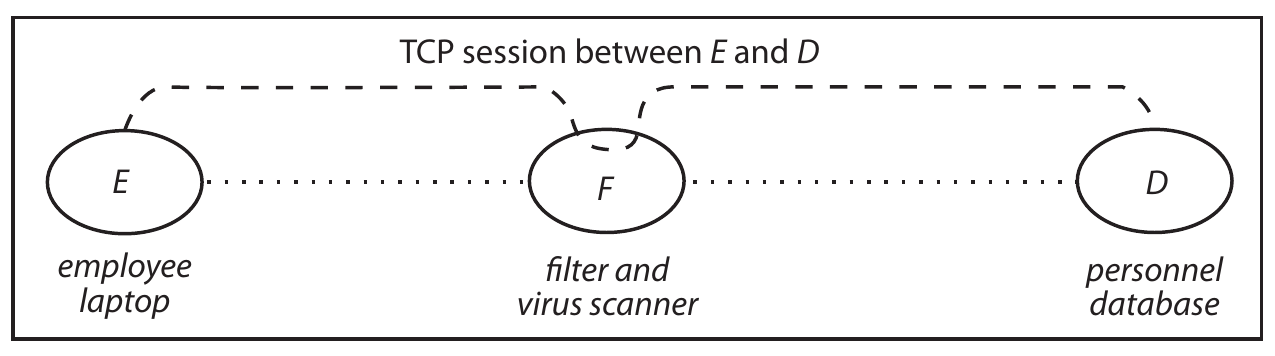}
\end{center}
\caption
{\small
{An enterprise IP network with security.  The dotted lines
represent paths of forwarders and links.}
\normalsize}
\label{fig:enterprise}
\end{figure}

As a consequence, layers in the new model are bigger than layers
in the ``classic'' Internet architecture of \cite{philo} and \cite{osi} 
(see Figure~\ref{fig:correspond}).
The advantage is that a bigger network/layer is more complete,
so its interface to an overlay network is the same as
the interface of an underlay network to it.
In other words, we have made networks composable like Lego bricks.

As demonstrated in \cite{cacm}, the compositional architecture
provides precise and comprehensive descriptions of how the Internet
works today.
For example, we see multiple layered IP networks,
each with its own purpose and geographical or logical span.
Each has its own namespace and routing, appropriate to its own
level of abstraction.
If the purpose of a network is narrow, one or more of its parts
might be vestigial, which causes no problems.

\section{Network services and properties}
\label{sec:Network-services-and-properties}

In this section, we use a security-oriented example to illustrate
the compositional model and sketch out how higher-level service
properties can be specified.
Figure~\ref{fig:enterprise} shows an enterprise IP network with
security features.
The employee's laptop has an IP address $E$ in the block reserved for
the personnel department.
The filter $F$ allows only machines belonging to the personnel 
department to access the personnel database.
It also does anti-virus scanning.
Because access from $E$ is allowed, there is now a TCP session between
$E$ and $D$.
Network security is intended to enforce two properties:
{\it P1:} All accesses to the personnel database are TCP sessions
initiated by user machines belonging to the personnel department, and
are free of known viruses.
{\it P2:} Packet streams traveling on the network paths cannot be read or 
tampered with by any untrusted machine.

Even though this example is simple so far, 
and the enforcement of {\it P2} is not
yet shown, a rigorous argument for {\it P1} requires these lemmas:
{\it L1a:} Every network path with the personnel database as destination
has a filter in it.
{\it L1b:} All 
packets of a TCP session go through the same filter instance,
so the filter can reconstruct the TCP byte stream to look for
viruses.
{\it L1c:} Addresses in the personnel block are only put into the
source fields of packets
by machines belonging to the personnel department.
{\it L1d:} Forwarders in network paths do not alter the source or
destination fields of packets
(necessary because {\it L1a} and {\it L1c} depend on these
fields).

\begin{figure*}[hbt]
\begin{center}
\includegraphics[scale=0.60]{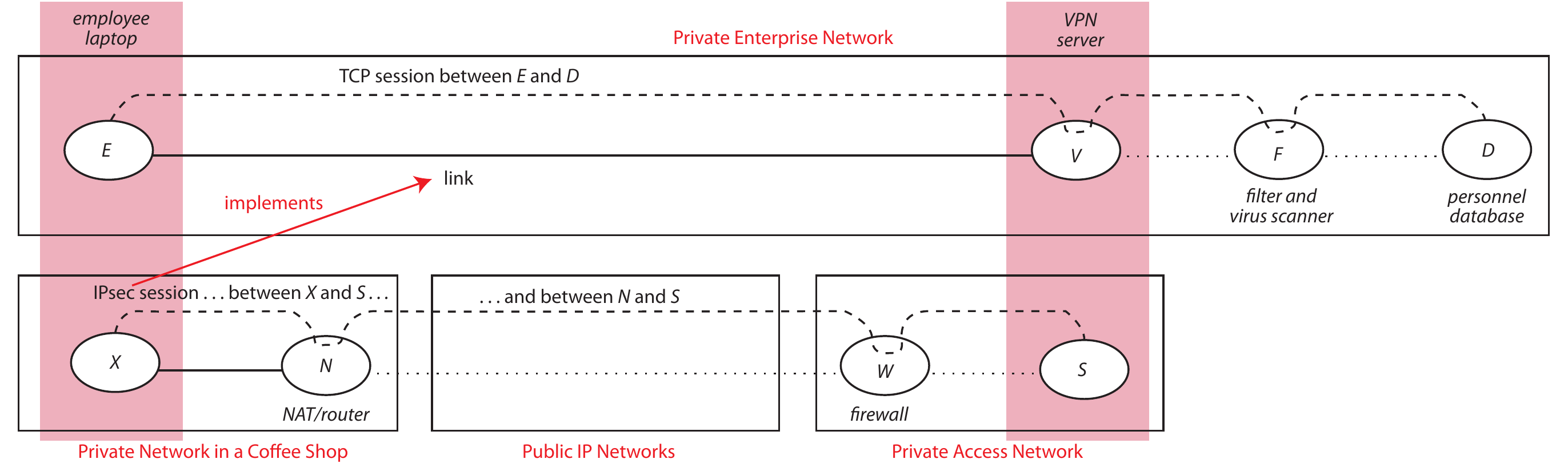}
\end{center}
\caption
{\small
{The VPN implementation of the enterprise network.}
\normalsize}
\label{fig:layering}
\end{figure*}

In today's environment, employees expect to be able to work on their
laptops wherever they happen to be.
So a more complete version of the enterprise network and its implementation
is shown in Figure~\ref{fig:layering}.
This figure shows that the enterprise network uses virtual private
network (VPN) technology to allow secure employee access from
remote locations.
The laptop is connected to two IP networks, one of which is 
layered on top of the other.
The leftmost pink bar represents the laptop, and the ovals represent
its members of (interfaces to) the two IP networks.
In the enterprise network it has IP address $E$, while in the IP
network of the coffee shop where the employee is working, it has
the short-term IP address $X$.
Network members $E$ and $X$ communicate through the operating system
of the laptop.

In this scenario, $E$ has created a virtual link, in the
enterprise network, directly to a VPN server.
This overlay link is implemented by an IPsec/ESP session in the
underlay, consisting
of bridged public and private IP networks.
TCP packets sent virtually on the enterprise
link are actually encrypted (headers included),
encapsulated in packets of the IPsec session, and transmitted through
the underlay IP networks.
The VPN server also authenticates the employee laptop,
checking that it has the private key of
enterprise identity $E$.

Throughout this example, for simplicity, we assume that all
middleboxes such as $V$, $F$, and $W$ are correct and trustworthy
in essential respects.
This will enable us to focus on the networks themselves.
So authentication in $V$ satisfies {\it L1c}, as the VPN server is only
one hop away from the laptop.

Now we consider security property {\it P2} of the enterprise network.
Because the trustworthiness of middleboxes is assumed and the 
necessary properties of forwarders are 
covered by {\it L1a, L1b,} and {\it L1d}, there is only one new lemma.
{\it L2:} All links
of the enterprise network are {\it secure},
in the sense that no untrusted machine
can read the packets on a secure link
(neither header nor data), and no machine can insert or alter packets
on a secure link.
In the figure, links to the right of $V$ are located
in enterprise buildings and secured physically.
The virtual link between $E$ and $V$ is secure if
every packet received by either endpoint is the data part of a
packet received in the IPsec session between $X$ and $S$.
The NAT/router alters the headers of IPsec packets, but not the headers
of encapsulated enterprise packets.

Finally, we add security property {\it P3}: machines at the edge
of the enterprise network are protected against flooding attacks.
This property is enforced by the access network in 
Figure~\ref{fig:layering},
where firewalls usually 
allow incoming packets to public IP address $S$, but begin to filter
out suspicious ones when there is a sudden surge in traffic.
Lemma {\it L3} states: All paths through the access network to machines
of the enterprise network pass through a firewall.

\section{Modular reasoning within the architecture}
\label{sec:Modular-reasoning-within-the-architecture}

This section, while informal,
is based on our formal semantics of the compositional
network architecture. 
It is written in Alloy \cite{alloy-book},
which is a blend of first-order predicate logic and
relational algebra. 
The Alloy Analyzer verifies properties automatically
for models of bounded size, making it an excellent tool for
experimentation with formal models.

\subsection{Verification of forwarding}
\label{sec:Verification-of-forwarding}

Existing tools for data-plane verification \cite{netverify} take
the forwarding rules in network elements and compute from them 
reachability, non-reachability, middlebox insertion,
and other important path properties.
We propose to do exactly the same thing, except that 
in our implementation each network in a 
layered architecture has separate forwarding rules and other tables
(see \S\ref{sec:Prototype-implementation}), and
each layer can be analyzed independently.
This kind of modularity is not exploited in any current verification
tools \cite{netverify2}.

The first advantage of this approach is that a layer/network has
not only properties we need to prove, but the structure and constraints
we need to prove it.
Analysis of the enterprise network in 
\S\ref{sec:Network-services-and-properties} can
use the fixed
naming scheme that allocates a block of names for machines of
the personnel department, 
regardless of the fact that the employee's laptop will have a
different IP address in each coffee shop, and that the source name
in its packets changes from $X$ to $N$ as the packets pass through
the coffee shop's NAT.
Analysis of the enterprise network can identify TCP packets and
check that they go through a filter that reconstructs TCP byte streams,
regardless of the fact that on some physical
hops they look like IPsec packets
and the TCP headers are encrypted.
Analysis can use as an axiom that forwarding through the enterprise
network does not modify the user-chosen fields of packet headers,
which satisfies lemma {\it L1d}.

The second advantage of this approach is that modularization
makes verification more efficient.
Lemmas {\it L1a} and {\it L3} are essentially the same
property---all packets with certain destinations
must pass through a security middlebox---but
they are verified in different (and simpler) networks.
To quantify the potential benefits, consider the example
of \cite{simple}, in which layered separation of a policy-driven network
(forwarding to middlebox instances) from a destination-driven network
is found necessary to prevent combinatorial explosion in the number
of rules.
With no layering, the forwarding rules would not fit 
in the TCAM of forwarders.
This suggests that, in comparison to the number $n$ of forwarding
rules in a non-modularized network, 
the number of rules in one layer of a two-layer architecture might be
$\sqrt{n}$ 
rather than $n/2$.

We studied a variety of network services and properties, and found
that {\it all} of the necessary lemmas could be verified in a single
layer.
About half of the properties apply directly to the layer in which
they are implemented, such as the {\it L1} lemmas and {\it L3}
in this example.
The other half apply to compositions of networks (next subsection).

\subsection{Hierarchical reasoning over forwarding\\and session properties}
\label{sec:Hierarchical-reasoning}

A one-to-one correspondence between implementing
sessions and implemented links allows bottom-up propagation of
properties verified within individual networks: any property known
to be true of an implementing session is also true of the implemented
link.
For this reasoning to be credible, there must be a ``tight seam'' 
between the two that does not allow any leakage of packets in or out.
This ``seam'' is built into our architecture and
implementation, rather than
being constantly and casually re-invented---it can be
scrutinized {\it once} with great care.

This hierarchical reasoning has an important advantage that we have not
seen in other verification work---it can prove theorems
that rely on properties of routing/forwarding
{\it and} on properties of session protocols.
It can also prove theorems that rely on the composition of completely
different forms of automated reasoning.
For example, property {\it P2} of 
\S\ref{sec:Network-services-and-properties}  
is a property on paths through the enterprise network,
including forwarders, middleboxes, and links.
We have already discussed properties of paths, forwarders, and middleboxes,
but what about the links ({\it L2})?
Links outside the walls of the enterprise are secure
only because they are implemented by a cryptographic
session protocol, which can be verified using pure mathematics and
temporal-logic model checkers.

Note that TCP sessions in the enterprise network cannot be encrypted
because $F$ works on plaintext.
Packets on enterprise
links are always plaintext, even if the implementing sessions are
encrypting them.
This works because the session protocol at the lower level decrypts
the data before delivering it upward to be received on the virtual link.

In our study of a variety of network services and properties,
the other half of the properties are verified in a single underlay
network,
but they are verified {\it for the benefit of} overlays.
In reasoning about the overlay networks, the properties become
assumptions about the behavior of their links.

\subsection{Packet traceability}
\label{sec:Packet-traceability}

Our final example is based on a ``service
network'' that underlies multiple private customer networks,
as in \cite{cabernet}.
The service network offers wide-area connectivity along with
enhanced performance, reliability, security, and customized
communication services.
Consider a customer network that must abide by
HIPAA regulations on medical-patient privacy.
Even if the service network normally examines customer 
packet contents for optimization or security filtering, it cannot examine
the contents of packets carrying patient records.

So the service network must distinguish normal packets from patient
records---without looking into them---and treat the patient 
records differently.
Our architectural model makes this easy.
In the customer network, there can be a separate topology of virtual
links used for patient records only.
In the service network, each virtual link is implemented by a
uniquely identifiable session.
Session identifiers then give the service network a reliable
indicator of how each packet should be treated.

This example illustrates how the architectural model provides
perfect traceability of packets across
all layers of a network hierarchy.
Traceability supports reasoning about
many security properties involving packet provenance.
It also has other uses; for example,
it provides accountability for dynamic, measurable properties
such as performance metrics.

\section{Prototype implementation}
\label{sec:Prototype-implementation}

The first goal for our prototype was to implement the architectural
model's basic structures, especially layering---which must form a
``tight seam'' at the interface between networks.
We can now show that this scaffolding is comprehensible, reusable,
and efficient.
Our next goal, in progress, is to gain insight from the prototype
on how the scaffolding could work with control functions,
management functions, session protocols, and verification in an
integrated approach to network development.

\subsection{Implementation of the model}
\label{sec:Implementation-of-the-model}

In this subsection we describe the P4 code needed to implement 
the data plane, specifically making the following three points:
(i) The implementation is completely general, and works for any set
of layered networks.
(ii) The P4 code is so regular that most of it can be generated
automatically, greatly reducing costs and risks of error.
(iii) The general-purpose code is easy to optimize, so that specific
networks can be implemented efficiently.

\begin{figure}[hbt]
\begin{center}
\includegraphics[scale=0.50]{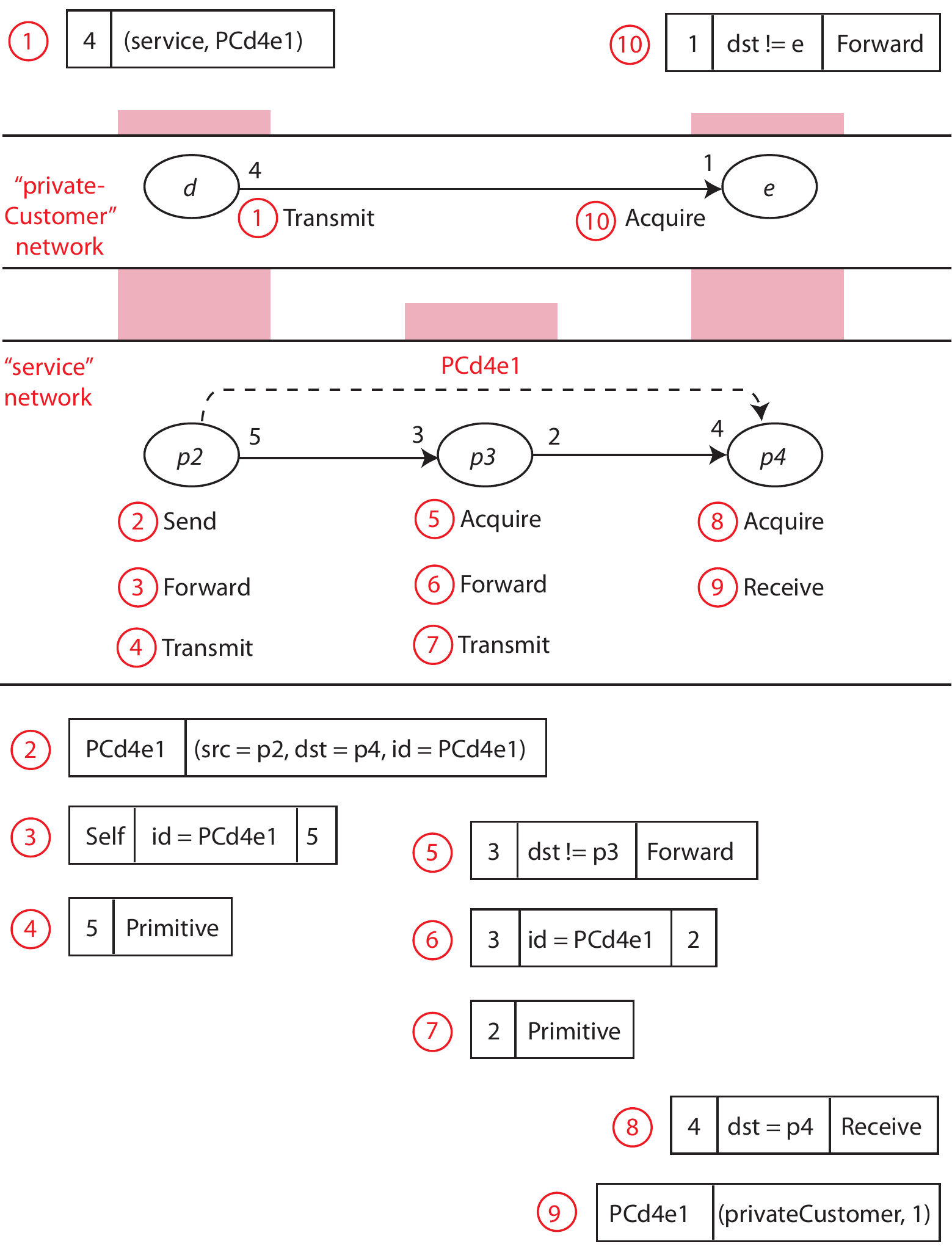}
\end{center}
\caption
{\small
{Transmission of a packet from {\it privateCustomer d} to
{\it e}, showing all the match-action functions applied.}
\normalsize}
\label{fig:functions}
\end{figure}

Figure~\ref{fig:functions} shows an example in which
a service network (from \S\ref{sec:Packet-traceability}) is implementing
the customer link from {\it d} to {\it e} with a session from
{\it p2} to {\it p4}, routed through service forwarder {\it p3}.
(The machine hosting {\it p3} does not host any members of the customer
network.)
The source-port field of packets in the service network acts as a
unique session identifier for multi-packet UDP sessions,
and the destination-port field acts
as a unique identifier of the customer network being served.

In unoptimized
P4 code, each network performs five functions, in each of which a
match-action table is applied.
The {\it Forward} function applies a forwarding table to determine
whether a packet should be dropped or sent on a specified outgoing
link.
{\it Send/Receive} applies when a network member sends/receives,
respectively, a packet in a session.
{\it Transmit/Acquire} applies when a network member sends/receives
a packet on a link.
All forwarding in the service network is based on session identifiers.
Figure~\ref{fig:functions} shows the functions and match-action table
entries applied as a packet in a customer network
is transmitted on a link from $d$ to $e$.

In Step 1, $d$ transmits a packet on the link with local identifier 4.
A Transmit table uses the link identifier as key, and lookup results
in either 
the literal {\it Primitive} (see Step 4)
or in an {\it external session,}
which is a {\it (network, sessIdent)} pair implementing the link.
The P4 function attaches {\it sessIdent = PCd4e1} 
to the packet as metadata, and next applies Send
for {\it p2} in the service network.

In Step 2, {\it p2} sends the packet in session {\it PCd4e1}.
The key in the Send table is the {\it sessIdent,} and
the action is an encoding of the session's header.
The program encapsulates the packet in this
header, adds the metadata {\it inLink = Self} to the packet, and
applies Forward for {\it p2} in the {\it service} network.

Step 3 applies the Forward table, with keys {\it inLink} and header
fields.
The result of a match is either the action {\it Drop} or an
{\it outLink} on which the packet should be transmitted.
The {\it outLink} is 5, and next (Step 4) the Transmit table is applied
with this key.
Step 4 in the service network is analogous to Step 1 in the customer
network.
Currently links in the service network are simulated as
physical ({\it Primitive}), and the packet is sent out the machine's
egress port 5.

In Step 5, as the packet is received on the other end of the link, 
it gets the metadata {\it inLink = 3}.
The keys into the Acquire table are the {\it inLink} and the
packet header, which is matched against header predicates in the table.
The possible actions in the table are {\it Receive} and
{\it Forward}.
Because {\it p3} is not the destination of the packet, the next
step will be to apply the Forward table.
Steps 6, 7, and 8 of the example are similar to Steps 3, 4, and 5.

In Step 8 the Acquire table says the packet will be received, and
in Step 9 the key in the Receive table is the session identifier
in the packet header.
The action in the table might be {\it Primitive}, meaning that this
session is being used by its machine for the machines's own purposes,
so the packet should simply be delivered to the operating system.
In this case the action is an {\it external link}, 
which is a {\it (network, linkIdent)} pair.
This means that the session is implementing the identified virtual link
in the external network, so the next step is to apply the
Acquire table for $e$ in the customer network, after stripping off
its service-network header.

With these five functions and suitable match-action tables, any set
of composed networks, no matter how complex their layering relationships,
can be implemented.
The P4 functions for a particular network depend chiefly on its header
format, so for known network types, most of the P4 code for a
set of composed networks can be generated automatically.

At the same time,
with knowledge of a specific network, it is easy to make optimizations
(which can be factored into code generation).
For example, we know that in this example all customer links are
implemented by sessions in the service network (as opposed to being
physical links), and all service links are
simulated physical links.
With this simplifying knowledge alone, 
a packet can be processed and forwarded at a machine, through two
layered networks, in a total of 4 match-action stages.
In an unoptimized program, there would be 8 stages.
On the Barefoot Tofino, our optimized prototype consumes about 3\%
of the compute resources, {\it e.g.,} ALUs, metadata, action buses,
and gateways.

The P4 functions for a network can have additional code for
network management, and this code is easy to place.
In the service network, {\it e.g.,} code for monitoring the performance
of a customer link would be plugged into Send and Receive for the
service session implementing the link.

\subsection{Integrated network development}
\label{sec:Integrated-network-development}

We have shown that each network is a module with its own separate set of
match-action tables and its own partition of the total set of
properties to be proved.
We have also shown that properties of different networks often take
a similar form.
This means that cumulative
experience will make it much easier to identify the
goals, constraints, and properties of specialized networks.
This is especially important for security goals, because it is
the modeling gaps that are exploited by adversaries.

The purpose of our formal semantics is to translate match-action
tables into network-wide behavioral properties.
So the formal semantics enables us to look at a network's
tables and say, {\it e.g.,} ``all paths to destination $d$ pass
through a firewall'' or ``consist of secure links.''

For small, static networks we can use the Alloy Analyzer to check
or generate match-action tables.
For real networks, in which tables are large and dynamic,
there must be control functions to generate and update them.
For our prototype we plan to experiment with a centralized controller.
With a specification of network topology and properties,
it may even be possible to generate controller code that is verified
correct, so that the tables it produces need not be verified.
This is already conceivable with tools such as Rosette \cite{Rosette},
and the modularity of networks may make it feasible.
Without network properties introduced as design principles, 
not only would controller generation be impossible, but the properties
necessary for data-plane verification would be missing.

\section{Related work}
\label{sec:Related-work}

Research on ``future Internet architectures'' has produced plans for
a number of clean-slate architectures, {\it e.g.,}
\cite{aip,mobilityFirst,RINA,ndn14}.
Each has its own special emphasis---so it is doubtful that any one of them
could meet all of the Internet's future needs, and they are not
compatible enough to merge into one unified design.
Compositional network architecture is not a clean-slate approach,
but rather a structured and modular way of describing networks as they
already are.
It easily covers special-purpose networks such as Named Data Networks
\cite{ndn14},
showing how they can be composed with other networks in a flexible
architecture.

The remaining related work concerns network verification.
Tools for verification of routing and forwarding, {\it e.g.,}
\cite{headerspace2,headerspace,veriflow,Anteater}, were 
mentioned in
\S\ref{sec:Verification-of-forwarding}.
These tools have the advantage of applying to existing networks.
They are equally applicable to networks with our architecture,
however, and modularity should make them even more scalable.
The Tiros verification tool 
\cite{AWS-tiros} already benefits from incorporating some
customer-level abstractions, albeit without the unifying compositional
model.

There has been much recent research activity on automated
verification of P4 programs, such as
ASSERT-P4 \cite{assert-p4},
the p4v tool \cite{p4v},
and Vera \cite{Vera}.
Generally speaking, these verification efforts focus on low-level,
service-independent properties of P4 programs, {\it e.g.,} is a
referenced header field valid?  Are array accesses in bounds?
Are there recirculation loops?
These should be very useful for making our P4 code
robust.

Not surprisingly, the most significant issue for these verifiers
is how to check higher-level, network-wide properties
without knowing the match-action tables that will be driving the
programs.
Sample tables can be given to a verifier, although this makes analysis a
hybrid of verification and testing.
ASSERT-P4 and Vera are both based on symbolic execution, and use
symbolic (partial) representations of table contents.
The p4v tool, on the other hand, comes with a language for specifying
the properties of match-action tables, so that verification can rely
on these properties if they are known.
This is the best solution, but useless without the table properties.
The value of our approach is that the properties of tables
will be readily available, as well as the specification of network-wide
requirements.

\section{Research plans}
\label{sec:Research-plans}

Due to the breadth and novelty of our approach, our current work
is exploratory in nature.
Completing the prototype will answer remaining feasibility
questions, including:
(i) How much changes when networks are dynamic rather than static?
(ii) How do we incorporate session protocols, especially those that
may require non-P4 custom code to perform encryption and decryption?
Then, we can proceed to
further research projects, with evaluation, to answer
more interesting questions such as these:
(iii)
Will easily-automated optimizations be sufficient to produce acceptably
efficient P4 code?
(iv) How much of the service design space can be covered by model-based
development?
(v) How much code generation and automated verification can actually
be achieved?
(vi) Can we evaluate architectural trade-offs in two dimensions---both
for implementation efficiency and for effectiveness of specification
and verification?

Inevitably, there will always be a need for custom code, in both
control and data planes.
One promising approach to making it trustworthy
is the development of verifiable programming
languages such as Dafny, which is general enough for
systems programming yet restricted enough for automated verification
\cite{ironfleet,dafnyColumn}.
Another promising approach is the development of reusable, customizable,
and verified middleboxes, for example on the Vigor platform \cite{vigor},
greatly reducing the need for writing unverified code.
Even middleboxes benefit from layering, because their functions within
separate networks can become separate modules.

In summary, this paper presents 
a constructive approach toward verifiable network services.
Of course it will not be sufficient, but there is ample
evidence to show that it is at least a necessary step forward.
It also has the potential to coalesce and magnify the benefits
of many other research projects on network programming, verification,
and security.
Although it may seem improbable, it is better than the impossible
of continuing forever to connect the world with untrustworthy network
services,
or the impossible of working bottom-up from unprincipled to principled
networks.

\bibliographystyle{abbrv}
\begin{small}
\bibliography{hotnets}
\end{small}

\end{document}